\documentclass[twocolumn,twocolappendix]{aastex701}

\newcommand\prg{\Pi}
\usepackage{bm}
\usepackage{CJK}
\usepackage{amsmath}
\usepackage{tabularx}
\usepackage{comment}

\begin{document}
\begin{CJK*}{UTF8}{gbsn}

\title{3D Radiation Magnetohydrodynamic Simulations of Gravitational Instability in AGN Disks: Challenges to Magnetic Stabilization in the Radiation-Dominated Regime} 

\author[orcid=0000-0003-3792-2888]{Yi-Xian Chen}
\affiliation{Center for Computational Astrophysics, Flatiron Institute, New York, NY 10010, USA}
\affiliation{Department of Astrophysical Sciences, Princeton University,  4 Ivy Lane, Princeton, NJ 08544, USA}
\email{yc9993@princeton.edu}
\author[orcid=0000-0002-2624-3399]{Yan-Fei Jiang}
\affiliation{Center for Computational Astrophysics, Flatiron Institute, New York, NY 10010, USA}
\email{yfjiang@flatironinstitute.org}
\author[orcid=0000-0002-6710-7748]{Jeremy Goodman}
\affiliation{Department of Astrophysical Sciences, Princeton University,  4 Ivy Lane, Princeton, NJ 08544, USA}
\email{jeremy@astro.princeton.edu}

\begin{abstract}
Gravitational instability (GI) is expected in the outskirts of AGN disks
with high accretion rates, which can lead to intense star formation
especially when radiation pressure is dominant. 
However, the role of magnetic fields in this process remains uncertain. 
Local isothermal and adiabatic simulations suggest that net vertical flux (NVF) 
can enhance magnetorotational instability (MRI) heating and amplify toroidal fields through dynamo action, 
potentially suppressing GI and fragmentation through heating and magnetic support.
We perform radiation magnetohydrodynamic (RMHD) shearing-box simulations with NVF corresponding to initial plasma $\beta_z=10^3$--$10^4$ and Toomre $Q_T\gtrsim1$ (or analogously $Q:= \Omega^2/2\pi G \rho_{\rm mid}\gtrsim 1$ where $\rho_{\rm mid}$ is the midplane density). 
In gas-pressure-dominated cases, 
MRI-driven turbulence generates substantial Maxwell stress, 
providing heating that dominates over gravito-turbulent heating and, 
when sufficiently strong, drives the disk to $Q>1$. 
In contrast, radiation-pressure-dominated cases remain susceptible to fragmentation despite MRI activity. 
Their high effective compressibility leads to 
large density fluctuations, 
allowing over-dense regions to collapse even in the presence of additional heating,
suggesting that stability may require mean $Q\gg1$ when the radiation pressure fraction is high.
With our moderate $\beta_z$, GI suppression arises primarily from heating rather than magnetic elevation: 
the midplane remains thermally supported and lacks the
coherent field to raise the disk to $Q\gg1$ through magnetic pressure.
We defer more strongly magnetized regimes with lower $\beta_z$ that might be capable of acheiving this to future RMHD simulations with more self-consistent treatments of wind-driven mass and energy loss. 
\end{abstract}


\section{Introduction} 

Disks around Active Galactic Nuclei (AGNs) are expected to be self-gravitating 
(Toomre $Q_T\lesssim 1$) beyond 
$\sim 10^3$ gravitational radii 
of the central supermassive Black Holes. 
Such disks regions are prone to fragmentation and star formation, 
unless 
gravito-turbulence can provide auxilary heating to the disk in addition to magneto-rotational instability (MRI) and support the disk at a $Q_T\approx 1$ quasi-steady state. 
In simulations of gas-pressure dominated disks, 
runaway fragmentation 
requires a dimensionless cooling timescale of $\tau_{\rm cool}\lesssim 3$ \citep{Gammie2001,JohnsonGammie2003}. 
However, 
\citet{Chen2023} performed radiation hydrodynamic simulations to show that in radiation pressure dominated environments, 
the disk is much more compressible and easier to fragment, which elevates critical $\tau_{\rm cool}$ to at least $\gtrsim 40$, 
implying star formation more intense than previously inferred from the criterion in the gas-pressure dominated limit. 
Such stars may provide power to the disk to maintain a $Q_T\sim 1$ quasi-steady state \citep{SirkoGoodman2003,TQM,ChenLiu2026}.

Recent global simulations of AGN accretion suggest the possibility 
of outer disk regions with highly magnetized midplanes 
$\beta\sim 1$ \citep{Hopkins2024,guo2025idealizedglobalmodelsaccretion}. 
Starting with zero net vertical flux (ZNVF) 
but sufficiently strong net toroidal fields, 
these disks can retain coherent, dynamically dominant magnetic fields capable of suppressing gravitational instability, 
rather than relaxing to classical ZNVF states with steady-state $\beta\sim 50$ and recurrent field reversals \citep{Hawley1995,Stone1996,Fromang2007,Davis2010,Simon2012}.
Whether strongly-magnetized states can be maintained 
without continued external magnetic flux injection remains debated \citep{Salvesen2016b,Squire2025,Chen2026,Tomar2026} since azimuthal flux is not conserved with open boundary conditions.
By contrast, 
it is better established that strong NVF $(\beta_z \lesssim 10^3)$ can both drive and sustain
a magnetically dominated state with $\beta \lesssim 1$, 
at least in
isothermal simulations \citep{Suzuki2009,BaiStone2013,Salvesen2016}. The local NVF setup
may therefore offer a more stringent (Maxwell) stress-test of the star-formation
scenario in AGN disks, even though the origin of strong NVF at large radii
may be uncertain. 

Under this context, 
\citet{Tsung2025} performed local NVF MHD simulations of
GI in application to AGN disks. 
They found that for an adiabatic gas ($\gamma = 5/3$),
initial vertical fields with $\beta_z \lesssim 10^3$ suffice to generate strong
toroidal fields through the MRI dynamo, yielding magnetically dominated midplanes
in quasi-steady state in spite of strong Newtonian cooling. 
Even with dimensionless cooling timescale of
$\tau_{\rm cool} \sim 1$, which would inevitably lead to fragmentation in the
absence of magnetic fields, 
the magnetically dominated disk sustains turbulence
with $\alpha_M \gtrsim 1$ and $\beta \lesssim 1$, maintaining both heating and pressure support against further collapse.
They also note that this stabilization 
requires sufficiently strong shear $q$ and a vigorous MRI
dynamo, 
unimpeded by resistivity. 
The outcome differs qualitatively in global
simulations of self-gravitating protoplanetary disks (PPDs), where local shear
can be reduced in overdense regions and non-ideal MHD effects suppress the
large-scale MRI dynamo \citep{Kubli2023}.

In this paper, we build on the GI-MHD simulations of \citet{Tsung2025} and the
GI-RHD simulations of \citet{Chen2023} to investigate the influence of
magnetic fields on gravito-turbulent AGN disks that are cooled self-consistently
by radiation. 
We consider both gas-pressure-dominated and
radiation-pressure-dominated regimes. 
In \S~\ref{sec:setup}, 
we describe the RMHD simulation setup,
including initial and boundary conditions, diagnostics, and
criteria for identifying bound objects. 
We present our numerical results in \S~\ref{set:SF_results}. Starting from
$Q_T \sim 1$ states with cooling timescales of $\sim 20\,\Omega^{-1}$, and
imposing moderate NVF with initial midplane plasma $\beta_z = 10^3$--$10^4$ to
drive MRI turbulence, we find that gas-pressure-dominated disks are  maintained at $Q\gtrsim 1$ with Maxwell stress being the dominant heat source, 
whereas radiation-pressure-dominated disks remain susceptible to
collapse. This difference arises from the higher effective compressibility of
radiation-dominated gas, which produces larger density fluctuations under
MRI-driven turbulence. In \S~\ref{sec:SF_discussion} we discuss the
implications of our results, 
together with the open questions and numerical
challenges for prospective studies. 
In particular, whether field amplification at 
lower $\beta_z$ can drive the disk into a persistently magnetically dominated
state that more robustly suppresses GI remains uncertain, and is deferred to
future work.



\section{Numerical Setup}
\label{sec:setup}
We use \texttt{Athena++} \citep{Stone2020} to solve radiation magneto-hydrodynamic (RMHD) equations 
in a three-dimensional Cartesian shearing box configuration \citep{Goldreich1965,Hawley1995} coupled with radiative transfer and self-gravity, 
similar to the configuration of \citet{Chen2023} with the additional inclusion of magnetic fields, or \citet{Chen2026} with the inclusion of self-gravity:

\begin{equation}
    \frac{\partial \rho}{\partial t}+\boldsymbol{\nabla} \cdot(\rho  {\boldsymbol v})=0
\end{equation}

\begin{equation}
\begin{aligned}
    \frac{\partial(\rho  {\boldsymbol v})}{\partial t}+\boldsymbol{\nabla} \cdot\left(\rho {\boldsymbol v} {\boldsymbol v} - {\boldsymbol B}{\boldsymbol B}+\mathbf{P}\right)=-\rho \boldsymbol{\nabla} (\Phi_{\rm sg}+\Phi_{\rm s})
    \\-2 \rho \Omega \hat{\mathbf{z}} \times \mathbf{v} - \boldsymbol{G}_{r}
    \end{aligned}
\end{equation}

\begin{equation}
\frac{\partial E}{\partial t} +\boldsymbol{\nabla} \cdot [(E+P_{\rm gas}) \boldsymbol{v}- {\boldsymbol B}({\boldsymbol B}\cdot{\boldsymbol v})]=-\rho \boldsymbol{v} \cdot \boldsymbol{\nabla} (\Phi_{\rm sg}+\Phi_{\rm s}) -cG^0_r
\end{equation}

\begin{equation}
    \frac{\partial \boldsymbol{B}}{\partial t}=\nabla \times(\boldsymbol{v} \times \boldsymbol{B})
\end{equation}

Here $\rho$ denotes gas density, $\boldsymbol{v}$ the
velocity, and $q = 3/2$ the shear parameter for Keplerian rotation. The tidal
potential of the shearing box is $\Phi_{\rm s} = \Omega^2 z^2/2 - q\Omega^2 x^2$.
We write the pressure in tensor form as $\mathbf{P} =
(P_{\rm gas}+B^2/2)\mathbf{1}$, with $P_{\rm gas}$ its scalar value and $\mathbf{1}$ the
identity tensor, and assume a mean molecular weight $\mu = 0.6\,m_p$,
appropriate for fully ionized gas of solar composition. The gas+magnetic energy density
$E = U_{\rm gas} + \rho v^2/2 + B^2/2$ comprises the internal energy
$U_{\rm gas} = P_{\rm gas}/(\gamma - 1)$ with $\gamma = 5/3$, the kinetic energy,
and the magnetic energy, where magnetic fields are expressed in units with
$4\pi\mu_0 = 1$. Vertical gravity from the central black hole enters the momentum
and energy equations as a source term through $\Phi_{\rm s}$. Radiation couples
to the gas through the space-like and time-like components of the radiation
four-force, $\mathbf{G}_r$ and $G_r^0$ \citep{Mihilas1984}, which we compute from
the frequency-integrated (gray) specific intensity evolved as a function of time,
position, and photon direction; implementation details are given in
\citet{Jiang2021}. For opacities, we take a constant electron-scattering value
$\kappa_s = 0.4~{\rm cm^2\,g^{-1}}$ together with small but finite Rosseland
and Planck absorption opacities, $\kappa_R = \kappa_P = 0.05~{\rm cm^2\,g^{-1}}$.
Realistic Planck opacities can exceed this value \citep{Jiang2020}, but the
choice is adequate provided the radiation and gas remain in thermal equilibrium,
as we find throughout the optically thick interior of the disk. Time integration
uses the second-order van Leer scheme, hydrodynamic fluxes are computed with the
HLLD Riemann solver, and both the intensity and the hydrodynamic variables are
reconstructed at second order. Finally, the disk self-gravitational potential
$\Phi_{\rm sg}$ is obtained by solving the Poisson equation

\begin{equation}
    \boldsymbol{\nabla}^{2} \Phi_{\rm sg}=4 \pi G \rho,
\end{equation}

using fast Fourier transforms \citep{Koyama2009,Kim2011}. 

\subsection{Initial Conditions}
\label{sec:initial}
The initial vertical profiles of a horizontally homogeneous disk characterized by surface density $\Sigma$, 
orbital frequency $\Omega$ and the midplane 
GI factor $Q$ (a proxy for Toomre parameter $Q_T$)
can be setup through the framework of \citet{Chen2023}, 
which we summarize here briefly. 
First, 
the midplane density is obtained from the definition of $Q$ 

\begin{equation}
    Q = \dfrac{\Omega^2}{2\pi G \rho_0},
\end{equation}

For an $n=3$ vertical polytrope profile with constant $P(z)/\rho(z)^{4/3}$ or constant radiation-to-gas pressure fraction, 
solving the hydrostatic equilibrium 
(an initially vertically homogeneous $B_z$ does not provide pressure support) 
and Poisson equation yields

\begin{equation}
\begin{aligned}
    \dfrac{z}{\sqrt{2 Q} H_0} =&
     \int_{0}^{\sqrt{1-\theta}}\frac{2 \sqrt{2}d w}{\sqrt{2 Q+1+\sum\limits_{i=1,2,3}\left(1-w^{2}\right)^{i}}},\\ \theta =& [\rho(z)/\rho_0]^{1/3}
\end{aligned}
\end{equation}

where the initial scale height can be expressed in terms of the midplane density and pressure $P_{\rm tot,0}$: 

\begin{equation}
    H_0 := \dfrac{\sqrt{P_{\rm tot,0}}}{\sqrt{\rho_0} \Omega } = \dfrac{\Sigma }{8\sqrt{Q} I_3(Q)  \rho_0} 
\end{equation}

such that $\int_{-\infty}^{\infty} \rho dz = \Sigma$. The integral is defined as

\begin{equation}
\begin{aligned}
I_{k}(Q) \equiv \int_{0}^{1} \frac{\left(1-w^{2}\right)^{k} d w}{\sqrt{2 Q+1+\sum\limits_{i=1,2,3}\left(1-w^{2}\right)^{i}}}.
\end{aligned}
\end{equation}

The result converges to $\Sigma\approx 2.5\rho_0 H_0$ at $Q = \infty$, which is consistent with a non-self-gravitating disk. 
The scale height defines a midplane sound speed $c_{s,\rm tot, 0} = H\Omega$ linked to the midplane temperature $T_0$ through the Eddington quartic

\begin{equation}
    P_{\rm tot,0} = \rho_0 c_{s,\rm tot, 0}^2 = \rho_0 \dfrac{\mathcal{R} T_0}{\mu} +\dfrac{a T_0^4}3
\end{equation}

and the vertical temperature profile is simply $ T(z)/T_0 = [\rho(z)/\rho_0]^{1/3}$.

One can also define $Q_T$ for a 3D disk through the vertically-averaged sound speed  according to:

\begin{equation}
    \left[c_{s,\rm tot}^2\right]_\rho \equiv \frac{\int_z c_{s,\rm tot}^2 \rho d z}{\int_z \rho d z},
\end{equation}
such that
\begin{equation}
    Q_T \equiv \frac{\left[c_{s,\rm tot}^2\right]_\rho^{1 / 2} \Omega}{\pi G \Sigma}.
\label{eqn:QTinitial}
\end{equation}

This $Q_T$, defined similarly to Equation 9 of \citet{Booth2019} or Equation 13 of \citet{2017MNRAS.471..317R}, 
has the advantage of capturing the vertically averaged properties and will smoothly connect to the global average quantity $Q_T$ in \S \ref{sec:sf_diagnostics}.
For a constant $\Pi$ vertical profile, the midplane $Q$ and the average $Q_T$ are linked through
\begin{equation}
    Q_T = \sqrt{\frac{ Q I_{4}(Q)}{16\left[I_{3}(Q)\right]^{3}} },
\end{equation}

With $\Sigma, \Omega$ and initial $Q_T$ specified, the vertical profiles in physical units are uniquely determined. 
In practice, we start from $Q_T = 1.6$ or $Q = 1.2$ (see table), 
such that we allow GI 
to gradually occur during an initial 
passive cooling/MRI phase and avoid strong initial transients.
We then set up initial vertical radiation flux by applying the radiative diffusion equation to 
$T(z)$ in the optically thick region and taper to a constant flux and energy density above the $\tau=1$ photosphere. 
The radiation field will adjust itself to smooth out its gradients near the photosphere after the simulation starts. 




To investigate the effect of magnetic fields,
We initialize with NVF

\begin{equation}
   B_z=B_0\left[1+\frac{1}{2} \sin \left(\frac{2 \pi x}{L_x}\right)\right],
\end{equation}

and $B_x = B_y = 0$. $L_x, L_y, L_z$ are the simulation domain dimensions. 
The characteristic $B_0$ is parameterized using 

\begin{equation}
    \beta_z = 2P_{\rm tot, 0}/B_0^2.
\end{equation}

In stratified MRI simulations with NVF, the mean plasma $\beta$ typically drops
substantially as the channel modes break down into turbulence. 
This drop is accompanied by dynamo amplification of the azimuthal magnetic field,  
which provides both
additional heating and magnetic pressure support, and can therefore significantly alter the
evolution of an otherwise gravitationally unstable, rapidly cooling disk. 
In a
razor-thin disk, linear analysis shows that the vertical field alone suppresses
Toomre instability if \citep{NakanoNakamura1978}

\begin{equation}
\frac{B_z^2}{\pi G \Sigma^2} \gtrsim 1,
\end{equation}

which corresponds to $\beta_z \lesssim 1$ for $Q \sim 1$. 
Realistic NVF usually lie below this threshold. 
The amplified toroidal field, however,
routinely reaches $\beta \lesssim 1$ in the saturated state of $\beta_z\lesssim 10^3$ NVF simulations
\citep{BaiStone2013,Salvesen2016,Tsung2025}. 
Although a stability analysis for
this configuration is considerably more involved \citep{kimostriker2001} and radiative diffusion introduces further complexities \citep{Thompson2008}, 
such azimuthal field may provide
enough magnetic support to stabilize the disk against Toomre instability,
independently of cooling. 


\subsection{Boundary Conditions}

We adopt the same boundary conditions, floors, and opacity tapering as
\citet{Chen2026}: 
shearing-periodic in $x$, periodic in $y$, and outflow
conditions in $z$ that prevent mass and energy injection into the domain
\citep{Jiang2013}. 
Floors are imposed at $10^{-2}\,T_0$ and
$10^{-4}\,\rho_0$, and the opacity in floor-dominated regions is reduced by a
factor $\min[1,\,10^{-2}(\rho/10^{-4}\rho_0)^2]$ so that the photosphere remains
inside the domain. 
The only addition is self-gravity, for which we apply vacuum
boundary conditions to $\Phi_{\rm sg}$ in $z$ \citep{Koyama2009}.

\subsection{Diagnostics}
\label{sec:sf_diagnostics}

We denote volume and horizontal averages by
\begin{equation}
    \langle X\rangle \equiv \frac{\int X\,dx\,dy\,dz}{\int dx\,dy\,dz}, \qquad
    \langle X\rangle(z) \equiv \frac{\int X\,dx\,dy}{\int dx\,dy},
\end{equation}
with $\langle X\rangle_{\rm mid} \equiv \langle X\rangle(z=0)$.
We also define the
density-weighted average
\begin{equation}
\langle X\rangle_\rho \equiv \frac{\int X\rho\,dx\,dy\,dz }{\int \rho\,dx\,dy\,dz}.
\end{equation}
The dimensionless stresses characterizing angular momentum transport and
heating are
\begin{equation}
\begin{aligned}
\alpha_i &= \frac{\langle W_i \rangle}{\langle P_{\rm tot} \rangle},
\qquad\mbox{in which}\\\qquad
R_{xy} &= \rho v_x \delta v_y, \quad
M_{xy} = -B_x B_y, \quad
G_{xy} = \frac{g_x g_y}{4\pi G},
\end{aligned}
\end{equation}
for $W_i \in \{R_{xy}, M_{xy}, G_{xy}\}$ (Reynolds, Maxwell, and gravitational),
where $\delta v_y = v_y + q\Omega x$ is the azimuthal velocity relative to the
background shear, and $\boldsymbol{g} = -\nabla\Phi_{\rm sg}$. We track
gravitational stability through the variables
\begin{equation}
    Q_T = \frac{\langle c_{s,\rm tot}^2\rangle_\rho^{1/2}\,\Omega}{\pi G \Sigma},
    \qquad
    Q = \frac{\Omega^2}{2\pi G \langle \rho \rangle_{\rm mid}},
\end{equation}
which reduce exactly to the initial values defined in \S~\ref{sec:initial} at
$t=0$.\footnote{Magnetic pressure is excluded from $Q_T$ because it does not
dominate support 
near the midplane in our simulations.} 
The thermal energy per
unit area is $U_{\rm 2D} = \langle U_{\rm tot}\rangle L_z$. 
For pressure ratios,
we average numerator and denominator separately rather than averaging the ratio
itself, which converges more reliably:
\begin{equation}
    \langle \Pi \rangle \equiv \frac{\langle P_{\rm rad}\rangle}{\langle P_{\rm gas}\rangle},
    \qquad
    \langle \beta \rangle \equiv \frac{2\langle P_{\rm tot}\rangle}{\langle B^2\rangle},
    \label{eqn:def_ratios}
\end{equation}
and analogously for the profiles $\langle \Pi \rangle(z)$ and
$\langle \beta \rangle(z)$ (see \citet{Chen2026} for details). Because $P_{\rm tot}$ falls off with height while
the initial $B_z$ does not, $\langle \beta \rangle(t=0) \sim 0.1\,\beta_z$
for our box sizes, 
with the low-pressure upper layers weighting the volume
average. 
By the end of each simulation, $B_y^2$ dominates the magnetic pressure. 

Following \citet{Gammie2001} and \citet{Chen2026}, we define the total energy
density including self-gravity,
\begin{equation}
    \mathcal{E} = U_{\rm tot} + \frac{\rho v^2}{2} + \frac{B^2}{2}
    + \rho\left(\Phi_{\rm s} + \dfrac{1}{2}\Phi_{\rm sg}\right),
\end{equation}

which evolves as

\begin{equation}
\begin{aligned}
    \dfrac{\partial \mathcal{E}}{\partial t} = -\nabla\cdot[(U_{\rm gas} + P_{\rm gas} + \dfrac{\rho v^2}2 + \rho (\Phi_{\rm s} + \Phi_{\rm sg})) \boldsymbol{v} \\+ \boldsymbol{B}\times (\boldsymbol{v}\times \boldsymbol{B})  + \boldsymbol{F}_r + \boldsymbol{F}_{\rm sg}], 
    \end{aligned}
\end{equation}
where $\boldsymbol{F}_r$ is the radiative flux and
$\boldsymbol{F}_{\rm sg} = (8\pi G)^{-1}\left(\Phi_{\rm sg}\nabla\dot{\Phi}_{\rm sg}
- \dot{\Phi}_{\rm sg}\nabla\Phi_{\rm sg}\right)$ is the gravitational energy flux
\citep{Jiang2013sg}. 
Integrated over the domain, energy conservation demands that the total dissipation rate equal the work done at the radial boundaries,
\begin{equation}
\begin{aligned}
    \langle Q^+\rangle &= \frac{3\Omega L_z}{2}\langle R_{xy} + M_{xy} + G_{xy}\rangle_{S_x}
    \approx \frac{3\Omega L_z\langle P_{\rm tot}\rangle}{2}\,\alpha,\\
    \qquad \alpha & \equiv \alpha_R + \alpha_M + \alpha_G,
    \end{aligned}
\end{equation}
where the approximation holds for states that are horizontally homogeneous on
average. Energy leaves through the vertical boundaries $S_z$ via the radiative,
Poynting, and advective fluxes per unit area,
$\langle F_r\rangle$, $\langle F_B\rangle$, and $\langle F_{\rm gas}\rangle$,
defined as in \citet{Chen2026}, with $\rho\Phi_{\rm sg}$ added to the advected
energy. Because our \texttt{Athena++} Poisson solver does not evolve the self-gravitational energy in
conservative form, $\boldsymbol{F}_{\rm sg}$ is not captured exactly and can
produce spurious energy changes above round-off (see discussions in \citet{Jiang2013sg,Mullen2021}). 
We
neglect its boundary contribution, which is small in the absence of significant
mass loss. 
Quasi-steady balance, $d\langle\mathcal{E}\rangle/dt \approx 0$,
then requires

\begin{equation}
    \alpha \approx \frac{2\left(\langle F_r\rangle + \langle F_B\rangle
    + \langle F_{\rm gas}\rangle\right)}{3\Omega L_z\langle P_{\rm tot}\rangle}.
    \label{eqn:normalize_flux}
\end{equation}

With $U_{\rm 2D} \sim L_z\langle P_{\rm tot}\rangle$,
this reduces to the classical
relation $\tau_{\rm cool} \equiv U_{\rm 2D}\Omega/\langle F_r\rangle \sim \alpha^{-1}$
\citep{Gammie2001} in
the absence of magnetic fields and outflows. However, 
other flux terms can also be non-negligible when strong outflows occur. Finally, time averages are denoted
$\langle X\rangle_t \equiv \int X\,dt / \int dt$.

\subsection{Determining bound objects}
\label{sec:bound}
To confirm fragmentation, 
we verify formation of gravitationally bound objects using the method of \citet{Mao2020,GRID}. 
For each local minimum in the gravitational potential field $\Phi$, 
we first identify the closed contour with the largest value of $\Phi =\Phi_{\rm max}$ which contains no other local mimima; this
region is the hierarchical binding parent (HBP). Within the HBP, we then find
the largest isocontour $\Phi' \leq \Phi_{\rm max}$ inside which
\begin{equation}
    \int_{\Phi \leq \Phi'} \left[ E_{\rm turb} + U_{\rm tot} + \frac{B^2}{2}
    + \rho\left(\Phi - \Phi'\right) \right] dV < 0,
\end{equation}
where $E_{\rm turb}$ is the kinetic energy relative to the region's
center-of-mass velocity. The enclosed region is the hierarchical bound region
(HBR): its combined turbulent, thermal, and magnetic energy is insufficient to
carry the gas over the bounding contour into a neighboring potential well. If
the potential well is sufficiently deep, the entire HBP is bound and the HBR
coincides with it.

\section{Results}
\label{set:SF_results}
\subsection{Choice of Parameters}
We present results from four RMHD simulations below,
two of which share a $(\Sigma, \Omega)$ combination corresponding to gas-pressure-dominated disks, 
while the other two lie in the radiation-pressure-dominated regime. 
These cases are selected to have $Q_T \sim 1$ with comparable, 
moderate cooling timescales of $\sim 20\,\Omega^{-1}$, 
enabling a controlled comparison of the effects of radiation pressure (see Figure 4 of \citet{Chen2023}). 
The initial parameters, 
along with the resulting $\prg$ and $\tau_{\rm cool}$, 
are summarized in Table~\ref{tab:para}. 
For turbulent viscosity parametrized by $\alpha$  \citep{Shakura1973}, these states correspond to instantaneous accretion rates of 

\begin{equation}
\begin{aligned}
\dot{M} \sim 0.2
\left(\frac{\alpha}{0.1}\right)
\left(\frac{\Sigma}{3\times10^5\,\mathrm{g\,cm^{-2}}}\right)^3
\\\left(\frac{\Omega}{4\times10^{-8}\,\mathrm{s^{-1}}}\right)^{-3}
M_\odot\,\mathrm{yr^{-1}},
\end{aligned}
\end{equation}

spanning sub- and super-Eddington accretion regimes for SMBHs with masses of $10^8$ to $10^9\,M_\odot$.

We adopt two values of the initial plasma beta, $\beta_z = 10^4$ and $10^3$, 
corresponding to moderately magnetized NVF configuration. 
The gas-pressure-dominated cases are performed at a resolution of $N_x \times N_y \times N_z = 128 \times 128 \times 512$ 
in a domain of size $L_x \times L_y \times L_z = 6H \times 6H \times 24H$ and are run towards quasi-steady states. 
Owing to the computational cost of full 3D RMHD simulations, we limit their duration to $t_{\rm end}=100\Omega^{-1}$. For the shorter, fragmenting radiation-pressure-dominated runs, we use a domain of $6H \times 6H \times 12H$ resolved by $128 \times 128 \times 256$ cells. These runs are terminated once substantial fragmentation develops, at times $t_{\rm end}$ reported in dimensionless form as $\tau_{\rm end}\equiv t_{\rm end}\Omega$ in Table \ref{tab:para}.


\begin{table*}
\centering
  \begin{tabular}{c|ccccccccc}
     \hline\hline
     Initial Parameters & $\Omega$ & $\Sigma$ & $Q_T$ & $T_0$  & $\rho_0$   & $\tau_{\rm cool,0}$  & $\Pi_0$ & $\beta_z$ & $\tau_{\rm end}$\\
     \hline
Unit &      [rad s$^{-1}$] & [g/cm$^2$] &  & [K] & [g/cm$^3$]  &  & &  \\
     \hline
        \texttt{S3e5O4e-9\_b4} & $4\times 10^{-9}$ & $3\times 10^5$ & 1.6 & 51705.4 &  $2.54\times 10^{-11}$  &  42.36 & 102.2 & $10^4$ &21 \\
        \texttt{S3e5O4e-9\_b3} & --& -- & -- & -- &  --  &  -- & -- & $10^3$ &24  \\
        \texttt{S3e5O4e-8\_b4} & $4\times 10^{-8}$ & -- & 1.2 & 28646.6 & $3.29\times 10^{-9}$  &  16.79 & 0.12 & $10^4$ &100\\
        \texttt{S3e5O4e-8\_b3} & -- & -- & -- &--  & -- &  --  & -- & $10^3$ & --\\
        \hline
        \hline
        \end{tabular}
   \caption{Summary of our simulation suite of marginally gravitationally stable disks. $\Omega, \Sigma, Q_T$ determines all other derived thermodynamic variables while $\beta_z$ sets the vertical magnetic field configuration. $\tau_{\rm cool,0}$ and $\tau_{\rm end}$ are the initial cooling time and the simulation end time normalized by $\Omega^{-1}$. -- means ``as above.''}
   \label{tab:para}
  \end{table*}

For the gas-pressure-dominated runs, 
we deliberately adopt 
a large vertical domain size $L_z$ to 
suppress strong outflows 
and maintain 
approximate mass conservation without introducing artificial mass rescaling \citep{Booth2019, BaiStone2013, Salvesen2016}.
However, 
when exploring more strongly magnetized cases with $\beta_z \lesssim 10^2$, 
we find that mass loss through outflows becomes significant even with a large $L_z$. 
These outflows are accompanied by significant advective energy fluxes that can dominate over radiative cooling in the system’s energy budget, 
thereby deviating strongly from the standard framework of GI study.
In such highly magnetized regimes, 
both mass and energy budgets (which could be essential to the outcome) 
become difficult to control, 
while adopting larger $L_z$ becomes  computationally inefficient for the current uniform-resolution setup. We therefore focus on cases with 
$\beta_z \gtrsim 10^3$. 

\begin{figure*}[htbp]
\centering
\includegraphics[width=1.0 \textwidth,clip=true]{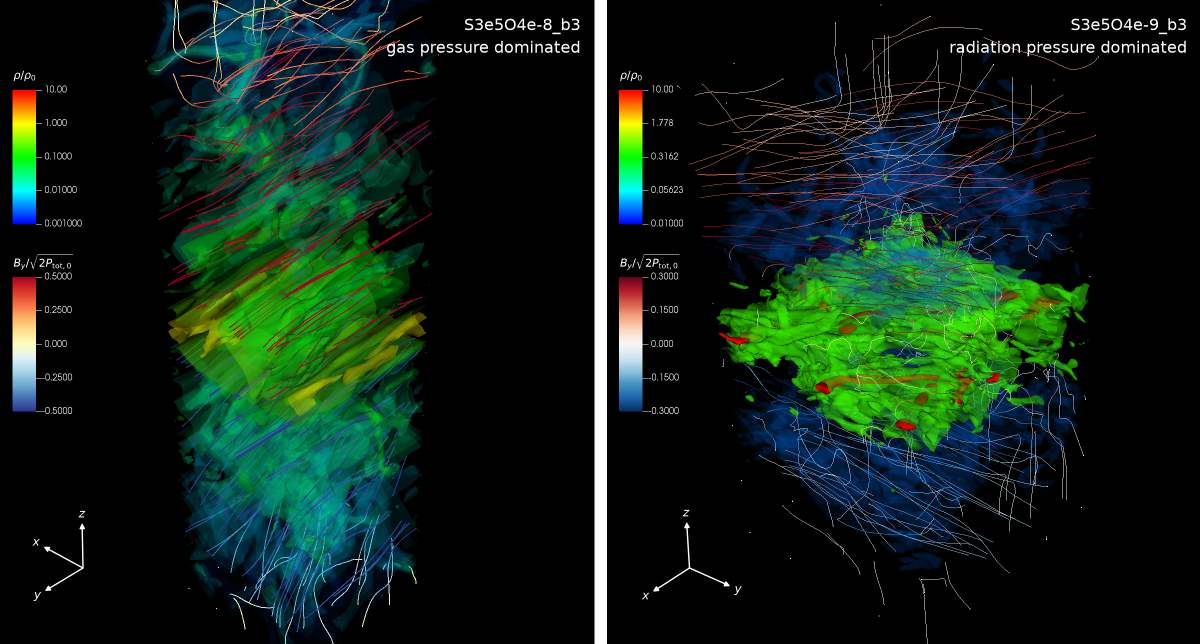}
\caption{3D renderings near the end of runs \texttt{S3e5O4e-8\_b3} (left) and \texttt{S3e5O4e-9\_b3} (right), 
illustrating a vertically extended, quasi-steady gas-pressure-dominated disk and a fragmenting radiation-pressure-dominated disk, respectively. 
Translucent isosurfaces show $\rho/\rho_0$, while magnetic field lines are colored by $B_y/\sqrt{2P_{\rm tot,0}}$, 
where $\rho_0$ and $P_{\rm tot,0}$ are the initial reference density and total pressure. Note that color scales and viewing orientations differ between panels. }
\label{fig:rendering}
\end{figure*}

Figure \ref{fig:rendering} shows representative snapshots near $t_{\rm end}$ for runs \texttt{S3e5O4e-8\_b3} (left) and \texttt{S3e5O4e-9\_b3} (right). 
The gas-pressure-dominated run settles into a vertically extended, 
quasi-steady state, 
whereas the radiation-pressure-dominated run undergoes fragmentation. 
The isosurfaces depict the normalized density $\rho/\rho_0$, 
while the streamlines trace the magnetic field (with magnitude dominated by $B_y$ despite starting at vertical at $t=0$) and are colored by the its azimuthal component, $B_y/\sqrt{2P_{\rm tot,0}}$. 
In the rest of this section, we examine the evolution of these two scenarios in detail.

\subsection{Gas-pressure Dominated, Quasi-steady States}

\begin{figure*}[htbp]
\centering
\includegraphics[width=1.0\textwidth,clip=true]{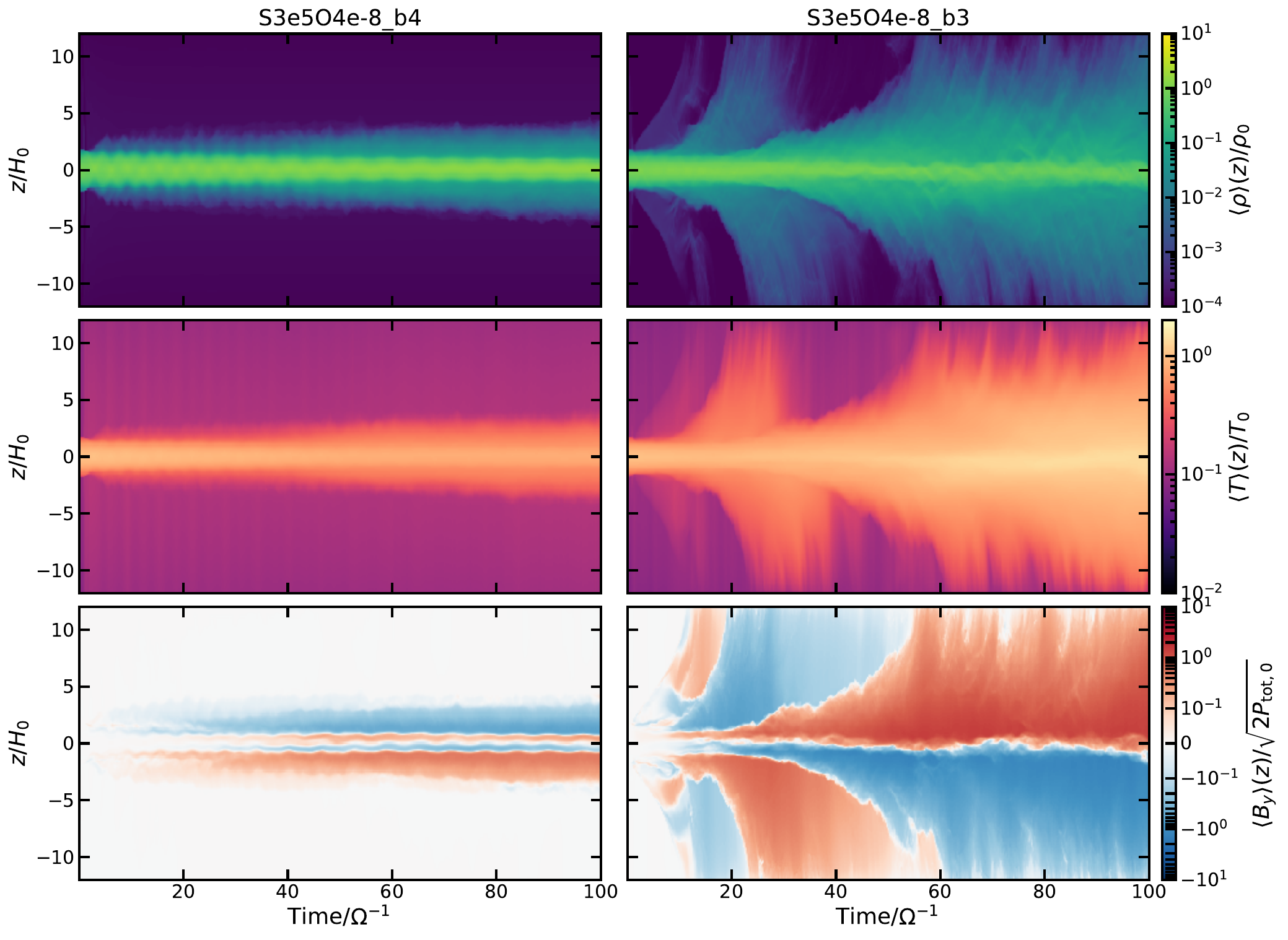}
\caption{Spacetime plots for run \texttt{S3e5O4e-8\_b4} (left) and \texttt{S3e5O4e-8\_b3} (right). From top to bottom, we show density, temperature, and azimuthal magnetic field as functions of height and time. 
Both runs evolve toward quasi-steady states without fragmentation, 
maintaining midplane layer with moderate density.}
\label{fig:spacetime_gasdominatedRMHD}
\end{figure*}

\begin{figure*}[htbp]
\centering
\includegraphics[width=1.0\textwidth,clip=true]{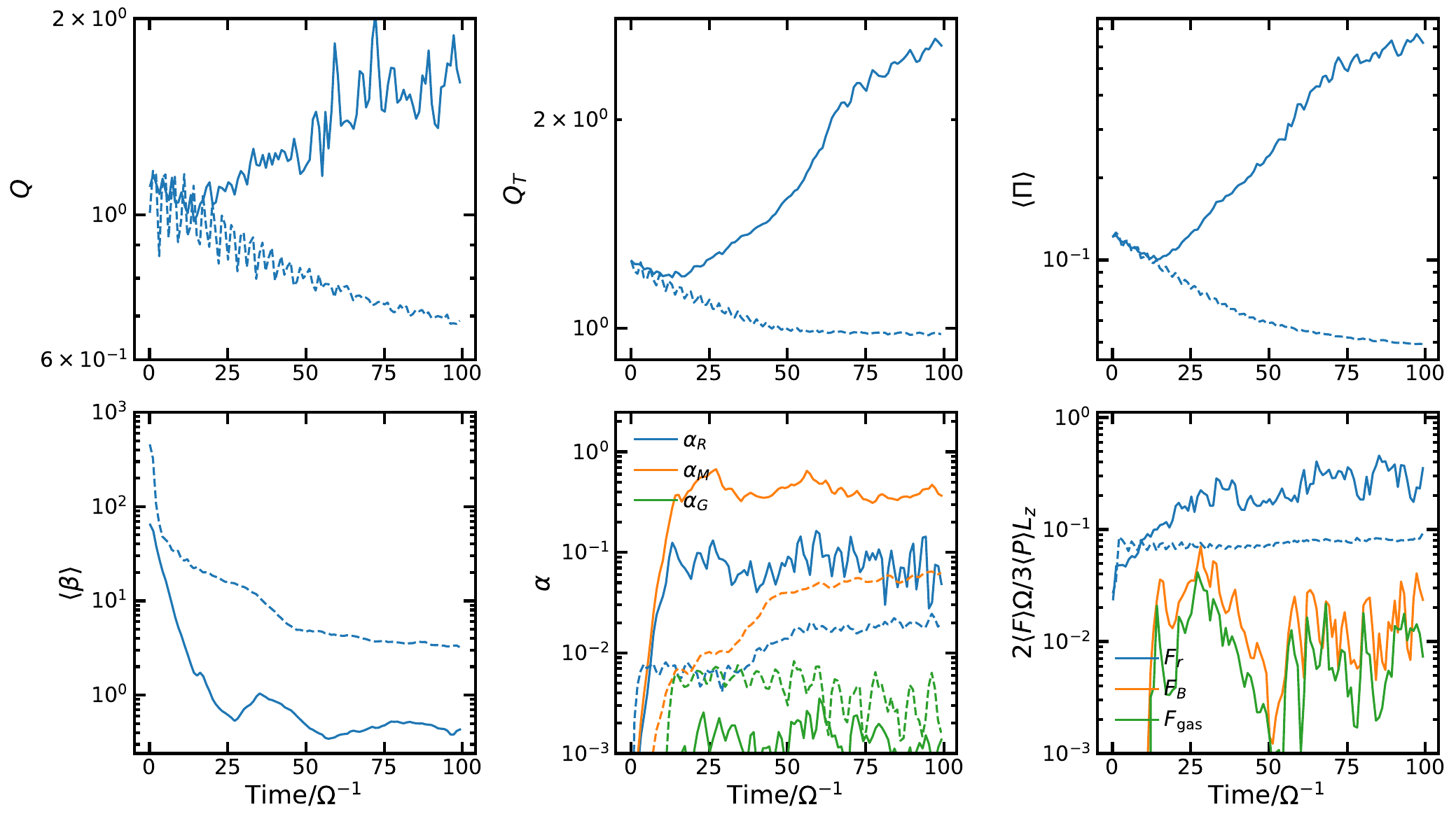}
\caption{The evolution of midplane and globally averaged statistics for run \texttt{S3e5O4e-8\_b4} (dashed lines) and \texttt{S3e5O4e-8\_b3} (solid lines). These average variables are defined in \S \ref{sec:sf_diagnostics}.}
\label{fig:stat_gasdominatedRMHD}
\end{figure*}

We first present results from runs \texttt{S3e5O4e-8\_b4} and \texttt{S3e5O4e-8\_b3}. 
The time evolution of horizontally averaged density, 
temperature, 
and azimuthal magnetic field profiles is shown in Figure~\ref{fig:spacetime_gasdominatedRMHD}, 
while the corresponding volume- and time-averaged diagnostics are summarized in Figure~\ref{fig:stat_gasdominatedRMHD}.
Both runs avoid collapse 
despite having moderate cooling times of $\sim 20\Omega^{-1}$, 
and develop substantial azimuthal fields driven by the dynamo from imposed net vertical flux. 
In the more weakly magnetized case, 
the 
quasi-steady state has a moderate $\langle \alpha_M \rangle_t \approx 0.07$ dominating the stress and heating budget, 
compared to $\langle \alpha_R \rangle_t \approx 0.02$ and $\langle \alpha_G \rangle_t < 0.01$. 
Outflows are negligible, with $v_z \approx 0$ near the vertical boundaries, 
so energy loss is dominated by radiative flux; 
all other normalized fluxes (see Equation \ref{eqn:normalize_flux}) remain $\lesssim 10^{-4}$. Consequently, 
thermal equilibrium is achieved by balancing turbulent heating against a normalized radiative flux of $\sim 0.1$, 
maintaining $Q \approx Q_T \approx 1$. 
In this state, net cooling is effectively halted and GI remains weak but non-negligible.

\begin{figure}[htbp]
\centering
\includegraphics[width=0.45\textwidth,clip=true]{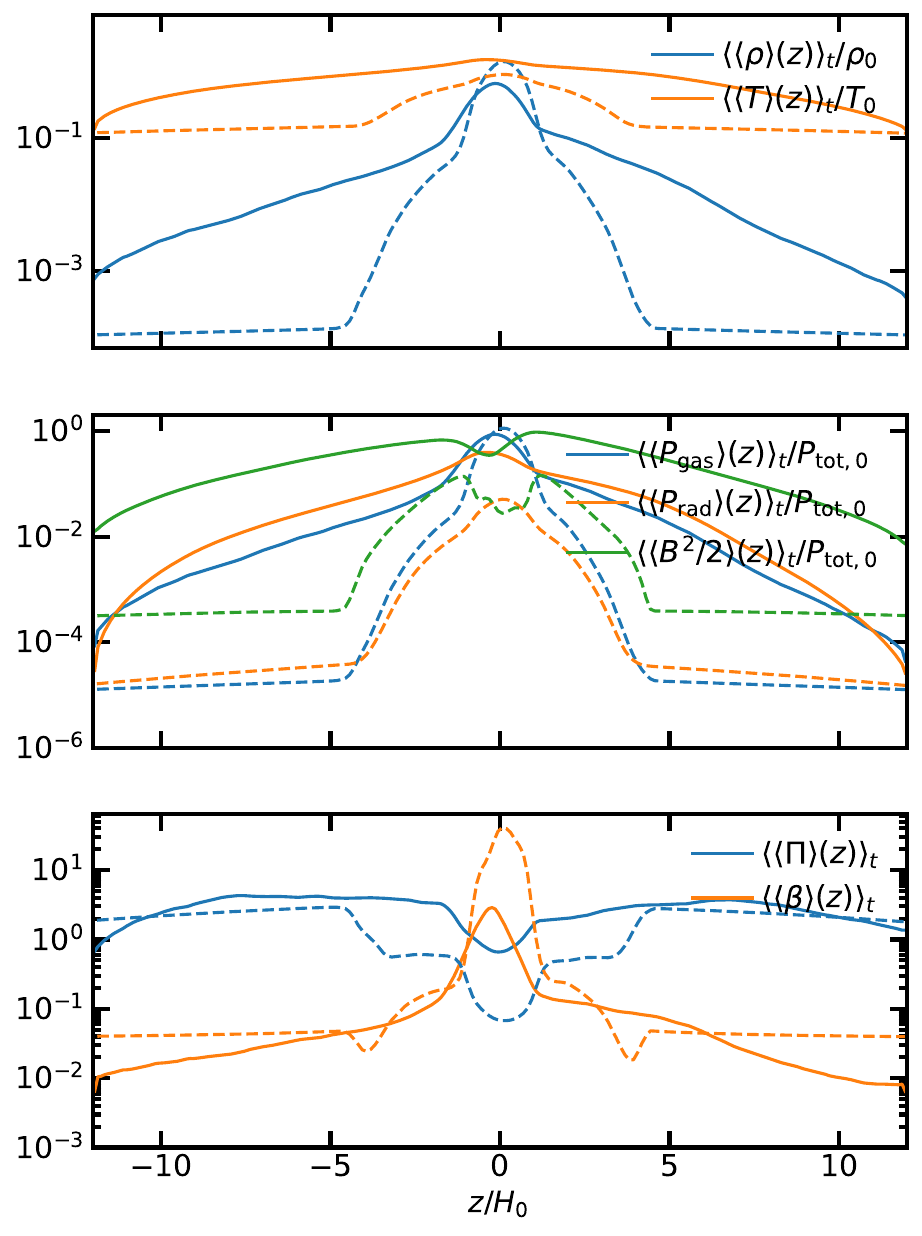}
\caption{The horizontally and temporally averaged vertical profiles for run \texttt{S3e5O4e-8\_b4} (dashed lines) and \texttt{S3e5O4e-8\_b3} (solid lines). }
\label{fig:vertical_gasdominatedRMHD}
\end{figure}

In the more strongly magnetized case, 
the azimuthal field exhibits large fluctuations but eventually $\langle \beta \rangle_t <  1$ on average. 
Nevertheless, 
the midplane remains compact $\rho_{\rm mid} \sim \rho_0$ and gas-pressure dominated, 
in contrast to the magnetically dominated high-altitude envelopes (see Figure~\ref{fig:vertical_gasdominatedRMHD}). 
The dominant heating source is magnetic stress, 
with 
$\alpha_M \sim 0.4$, 
significantly larger than in the weak-field case.
Outflows transport both magnetic and gas energy to high altitudes, 
making the Poynting flux non-negligible. 
However, 
the fraction reaching the vertical boundaries remains $<20\%$ of the radiative cooling rate in quasi-steady state, 
whereas the total cooling is subdominant compared to the strong net stress heating. 
As a result, 
the excess heating inflates the disk, 
leading to substantial increases in $Q$, $Q_T$, and the 
vertically averaged $\langle \Pi \rangle$, 
ultimately suppressing gravitational instability and choking $\alpha_G$. 
However, 
increase in temperature significantly enhances the cooling rate of a gas pressure dominated disk, 
so the high $\alpha$ disk 
shows tendency of settling to a steady state with stronger normalized cooling flux. 

By averaging over the final $\sim 20\,\Omega^{-1}$ for each run, 
we obtain representative vertical profiles shown in Figure~\ref{fig:vertical_gasdominatedRMHD}.
These profiles are consistent with the trends identified above: 
run \texttt{S3e5O4e-8\_b4} remains in a relatively compact, thermally supported state, whereas run \texttt{S3e5O4e-8\_b3} enters a $\beta \sim 1$ state where magnetic pressure is comparable to the thermal pressure. 
The latter case exhibits a more extended vertical structure, 
with lower density and $\beta$, 
and correspondingly higher temperature and radiation pressure fraction. 

Long-term isothermal simulations with net vertical flux
\citep{BaiStone2013,Salvesen2016} show stochastic dynamo activity: the mean
azimuthal field reverses sign quasi-periodically, producing the familiar
``butterfly'' pattern, although its period and morphology differ from those
in ZNVF simulations. Recurrent reversals originating at the midplane indicate
a dynamo that generates large-scale field in the turbulent midplane and
launches it vertically. 
Because of computational cost, our runs span only $\sim 100\,\Omega^{-1}$, too
short to sample multiple dynamo cycles or to assess the long-term evolution of
the field. 
In run \texttt{S3e5O4e-8\_b3}, for example, $B_y$ changes sign across
$z = 0$, 
so the azimuthal magnetic pressure 
contributes little to midplane support despite its dominance in the envelopes. Whether this antisymmetric configuration
persists, 
as in the steady states with midplane $B_y$ reversals found in recent global
simulations \citep{Zhang2026}, 
or whether continued dynamo action
eventually establishes a coherent $B_y$ that can magnetically support the
midplane, 
remains uncertain. 
We therefore focus on the role of MRI in quenching
GI rather than on the long-term dynamo behavior.


\subsection{Radiation-pressure-dominated, Fragmenting Cases}
\label{sec:radresults}

\begin{figure*}[htbp]
\centering
\includegraphics[width=1.0\textwidth,clip=true]{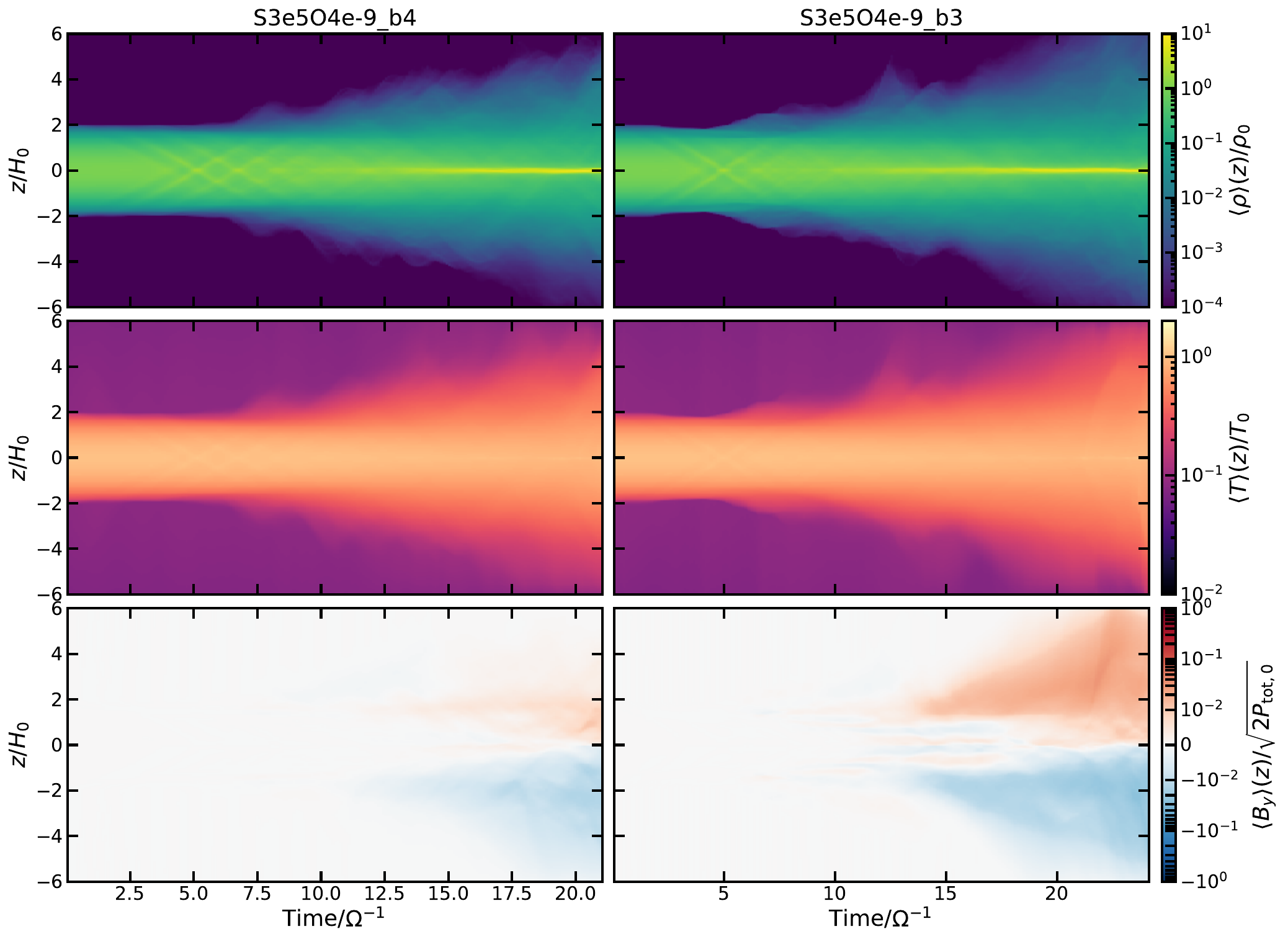}
\caption{Spacetime plots for run \texttt{S3e5O4e-9\_b4} (left) and \texttt{S3e5O4e-9\_b3} (right). The plotted quantities are similar to Figure \ref{fig:spacetime_gasdominatedRMHD}.}
\label{fig:spacetime_raddominatedRMHD}
\end{figure*}

\begin{figure*}[htbp]
\centering
\includegraphics[width=1.0\textwidth,clip=true]{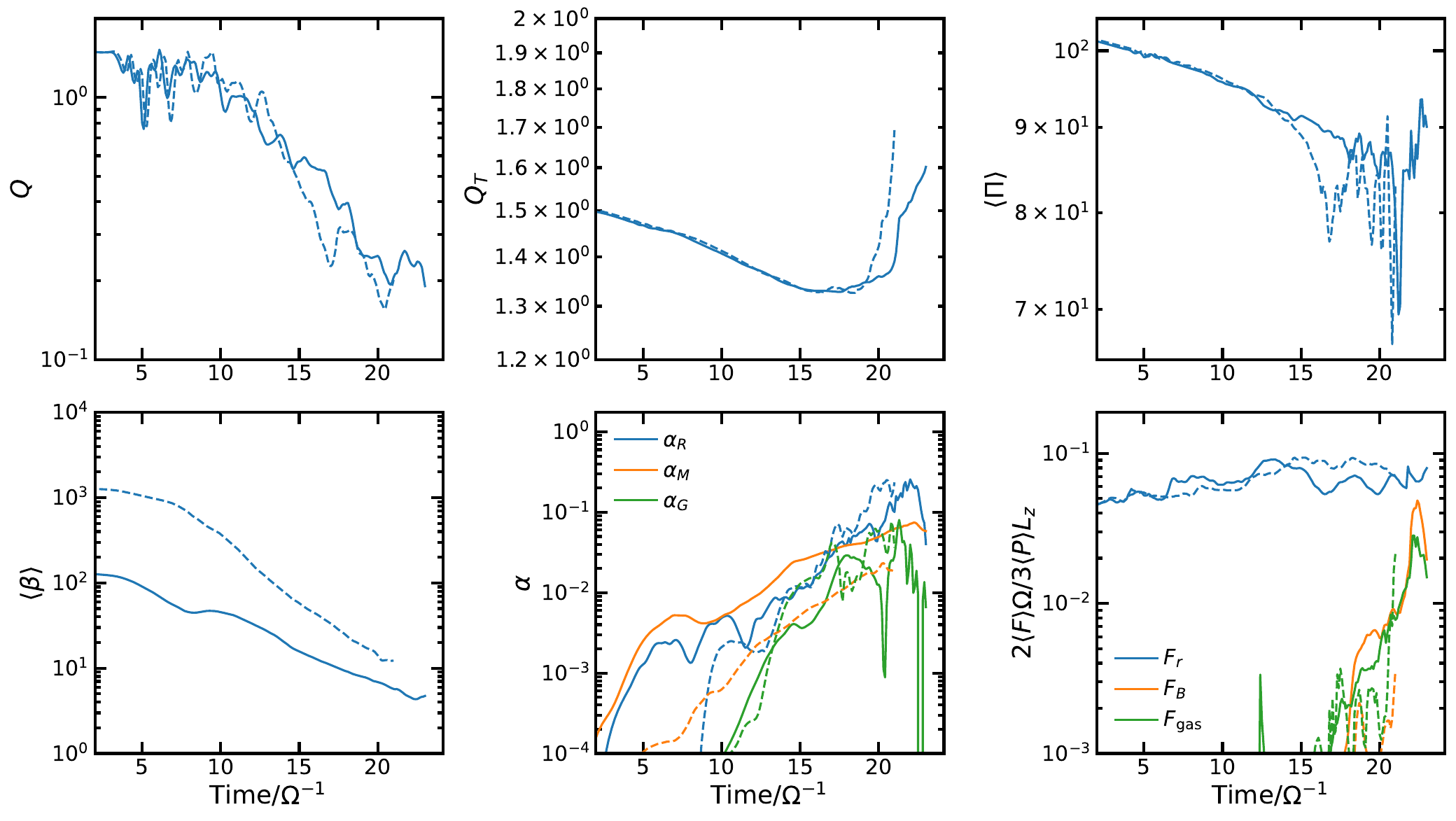}
\caption{The evolution of midplane and globally averaged statistics for run \texttt{S3e5O4e-9\_b4} (dashed lines) and \texttt{S3e5O4e-9\_b3} (solid lines).}
\label{fig:stat_raddominatedRMHD}
\end{figure*}

We next present results from runs \texttt{S3e5O4e-9\_b4} and \texttt{S3e5O4e-9\_b3}. The time evolution of horizontally averaged density, temperature, and azimuthal magnetic field profiles is shown in Figure~\ref{fig:spacetime_raddominatedRMHD}, with corresponding volume- and time-averaged diagnostics in Figure~\ref{fig:stat_raddominatedRMHD}.
In both cases, 
the disk undergoes fragmentation, 
becoming increasingly compressed toward the midplane. 
During the first few cooling timescales, net cooling drives declines in $Q$, $Q_T$, and $\Pi$. Fragmentation sets in at $t \sim 20\,\Omega^{-1}$, when $Q \sim 0.2$, leading to the formation of bound structures. 
At this stage, strong $\alpha_G$ and $\alpha_R$ are generated by chaotic motions near the midplane, 
temporarily increasing $Q_T$ and $\Pi$. However, this heating is intermittent, mainly heats up the upper layers of the disk and insufficient to halt collapse, 
and individual clumps continue to contract toward higher central densities until $\rho/\rho_0 \sim 10^4$.
In the moderately magnetized case \texttt{S3e5O4e-9\_b3}, 
Maxwell stress and Poynting 
flux become non-negligible, 
but remain subdominant and do not qualitatively alter the fragmentation outcome.

\begin{figure}[htbp]
\centering
\includegraphics[width=0.45\textwidth,clip=true]{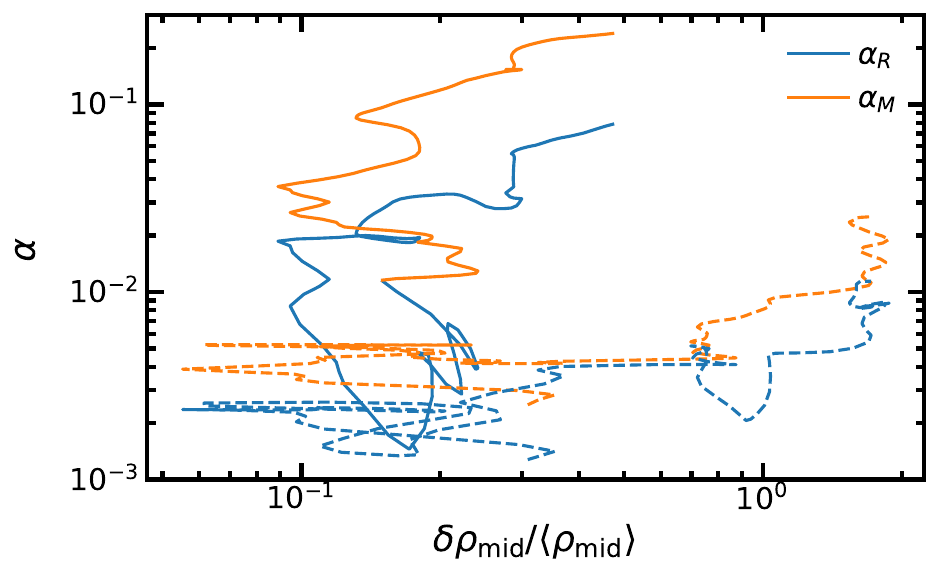}
\caption{Comparison of normalized midplane density fluctuations, 
$\delta\rho_{\rm mid}/\langle\rho\rangle_{\rm mid}$ from 5-15 $\Omega^{-1}$, versus stress parameter $\alpha$ for runs \texttt{S3e5O4e-8\_b3} (solid) 
and \texttt{S3e5O4e-9\_b3} (dashed), illustrating the larger density contrasts 
required to reach comparable stress levels in the radiation-pressure-dominated case.} 
\label{fig:density_dispersion}
\end{figure}

Compared to the gas-pressure-dominated runs, the most notable difference 
is the much slower initial growth of $\alpha_M$ and $\alpha_R$ in the radiation-pressure-dominated cases. 
For example, 
in the $\beta_z = 10^3$ runs, 
by $t \sim 15\Omega^{-1}$, prior to the onset of strong GI, the gas-pressure-dominated case has already reached MRI stresses of $\sim 0.1$, 
whereas the radiation-dominated case remains below $\sim 0.01$.
This behavior likely reflects higher compressibility of radiation-pressure-dominated gas: 
similar linear perturbations ($\delta \rho/\rho \sim \delta v/c_{s,\rm gas}$ when radiative diffusion is efficient) generate substantially weaker stresses in terms of the total pressure. As a consequence, 
(i) the associated turbulent heating is insufficient to balance cooling, 
and
(ii) reaching comparable stress levels requires much larger density fluctuations. 
These enhanced fluctuations inevitably trigger local collapse once sufficiently overdense regions form.
This trend is illustrated in Figure~\ref{fig:density_dispersion}, which shows the evolution of midplane 
(i.e., most fragmentation-relevant) 
density fluctuations up 
to $t \sim 15\Omega^{-1}$ for the $\beta_z = 10^3$ runs. 
The radiation-pressure-dominated case requires an order-of-magnitude 
larger density contrast to achieve similar 
$\alpha_R$ and $\alpha_M$, and therefore enters the nonlinear regime more rapidly.
This result can also be interpreted as an effective upper limit 
on $\alpha$ imposed by the compressibility of gas under influence of radiation pressure. 
A simple scaling follows from $\alpha_R \sim \rho\delta v^2 / P_{\rm tot}$ 
combined with the compressibility relation between $\delta v$ and $\delta \rho$, 
yielding $\alpha_R \sim (\delta \rho/\rho)^2/(1+\Pi)$, or equivalently $\delta \rho/\rho \sim \sqrt{\alpha_R(1+\Pi)}$ when gas and radiation pressure coexist \footnote{For GI-only turbulence the scaling may be more complicated since effective $\alpha$ is less scale-free}. 
Thus, achieving a given level of MRI-driven Reynolds stress 
in radiation-pressure-dominated disks necessarily entails much larger density contrasts, 
which in turn promote local collapse unless the system begins with substantially larger $Q$. 
Large density
fluctuations in radiation-dominated MRI turbulence are well documented in ZNVF
RMHD simulations, both unstratified \citep{Turner+2003} and stratified
\citep{Turner2004,Hirose2009,Blaes2011}, and have been confirmed in unstratified
RMHD runs with NVF \citep{Jiang2013sat}. 
Whether the same holds in
stratified NVF disks has not yet been examined in details, but we offer a preliminary confirmation of this feature in Appendix \ref{app:nsg}, where we simulate non-self-gravitating analog of run \texttt{S3e5O4e-9\_b3} to isolate the effect of gravitational collapse. 

\begin{figure}[htbp]
\centering
\includegraphics[width=0.45\textwidth,clip=true]{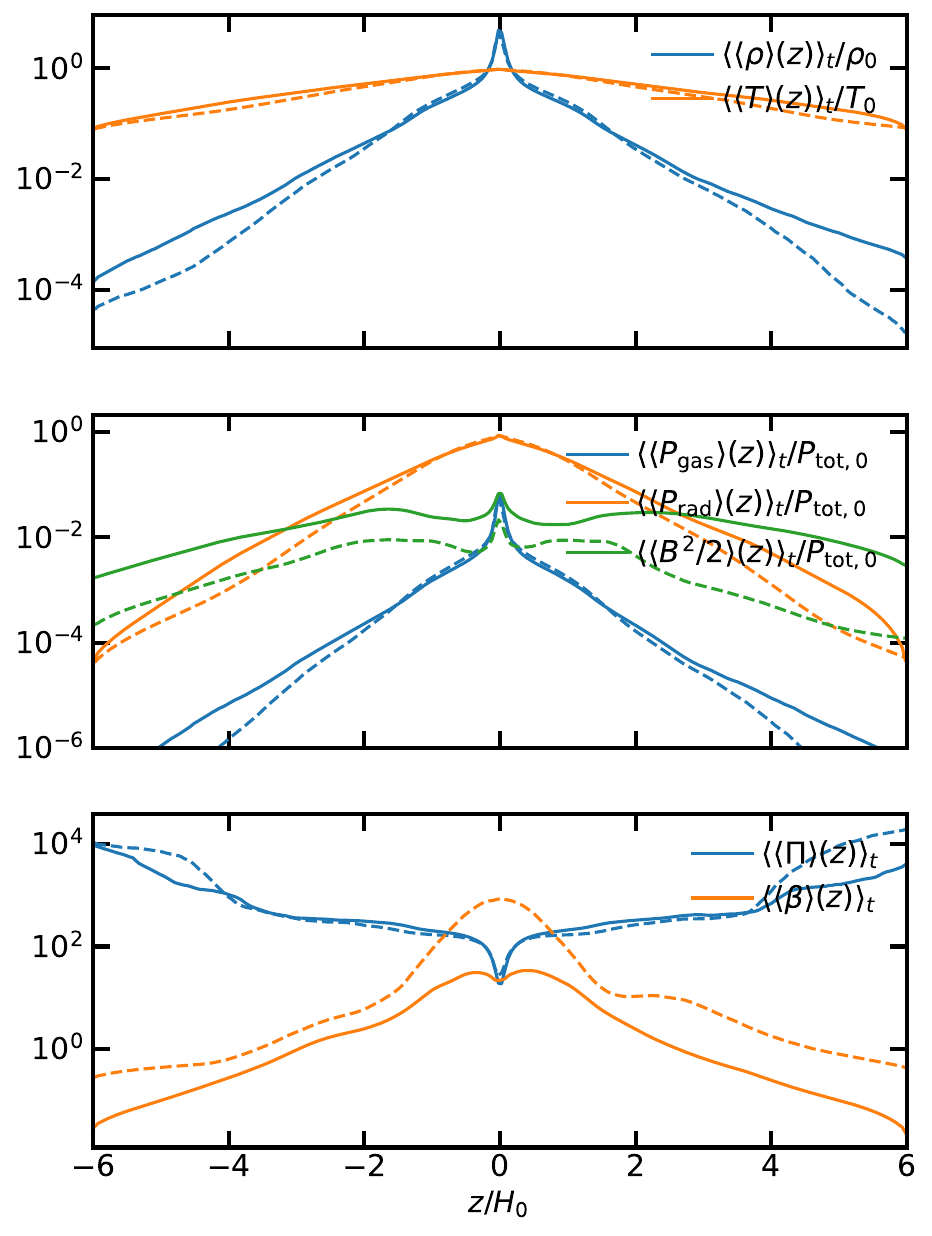}
\caption{The horizontally and temporally averaged vertical profiles for run \texttt{S3e5O4e-9\_b4} (dashed lines) and \texttt{S3e5O4e-9\_b3} (solid lines).}
\label{fig:vertical_raddominatedRMHD}
\end{figure}

Without loss of generality, we plot the average profiles of the radiation dominated runs in Figure \ref{fig:vertical_raddominatedRMHD}, over the final $10\Omega^{-1}$ of each simulation. 
In both cases, 
by the time of collapse the disk remains strongly radiation-pressure dominated, 
with a pronounced spike in density and pressure near $z=0$, 
while the temperature profile is relatively flat and close to isothermal.
Cooling in the dense midplane reduces the local radiation pressure fraction from $\Pi \sim 100$ to $\sim 10$, 
whereas the surrounding disk retains $\Pi$ comparable to or exceeding its initial value ($\sim 100$).
In the $\beta_z = 10^3$ case, 
the disk becomes marginally magnetically dominated near the surface layers, 
but this does not significantly influence the overall dynamics or alter the collapse behavior.

It is worth noting that we first performed lower-resolution, radiation-pressure-dominated runs with $Q_T(t=0)=1.2$, 
comparable to the gas-pressure-dominated cases, 
which led to even more rapid fragmentation (not shown). 
We then adopted a milder initial condition with $Q_T(t=0)=1.6$ at higher resolution, which again robustly fragments.
A natural question is whether starting from $Q_T(t=0) \gg 1$, so that
the MRI can fully develop and supply sustained heating before GI sets in, would
prevent fragmentation. 
This is plausible, but long-term thermal instability may
still drive the disk toward collapse and back to $Q_T \sim 1$ (runaway cooling and heating are both possible), 
as in the ZNVF
simulations of \citet{Jiang2013}. Appendix~\ref{app:nsg} therefore also serves a
second purpose: 
in the absence of self-gravity, our radiation-pressure-dominated 
runs with $\beta_z \sim 10^3$--$10^4$ do not undergo initial thermal collapse and, if
anything, expand. 
This supports attributing the collapse in our fiducial runs 
to self-gravity rather than to thermal instability, 
although it does not serve as a
dedicated study of thermal instability in NVF MRI, which remains largely unexplored.


Finally, the regime $\beta_z \lesssim 10^2$ remains uncertain. 
We have tested a $\beta_z = 10^2$ case for our radiation pressure dominant initial profiles 
and found clump formation 
accompanying strong outflows that remove half of the mass from the box within $20\Omega^{-1}$, 
but caution against over-interpreting its implication. 
Isothermal and adiabatic simulations
suggest 
they evolve toward fully magnetically dominated states \citep{BaiStone2013,Tsung2025}, 
but
the outcome may sensitively depend on radiative cooling and the treatment
of boundary conditions 
and mass-rescaling in presence of outflows. 
Moreover, even a disk with mean $\langle\beta\rangle \lesssim 1$ may not be
stable if it is radiation-pressure dominated. 
Its large density fluctuations
also implies a broad spread in local $\beta$: 
gas collecting along field lines
can form overdensities 
that are thermally rather than magnetically supported,
and these may still collapse locally. 
Our clumps show this tendency,
with magnetic pressure subdominant at their centers (see Figure~\ref{fig:clumps_raddominated}).
To probe more magnetized regimes, we again appeal to future dedicated RMHD simulations of MRI (both ZNVF or NVF) in stratified shearing boxes where
turbulence, outflows, dynamo action, and thermal stability 
are studied
independent of self-gravity. 

\subsection{Clump Properties}

\begin{figure*}[htbp]
\centering
\includegraphics[width=0.8\textwidth,clip=true]{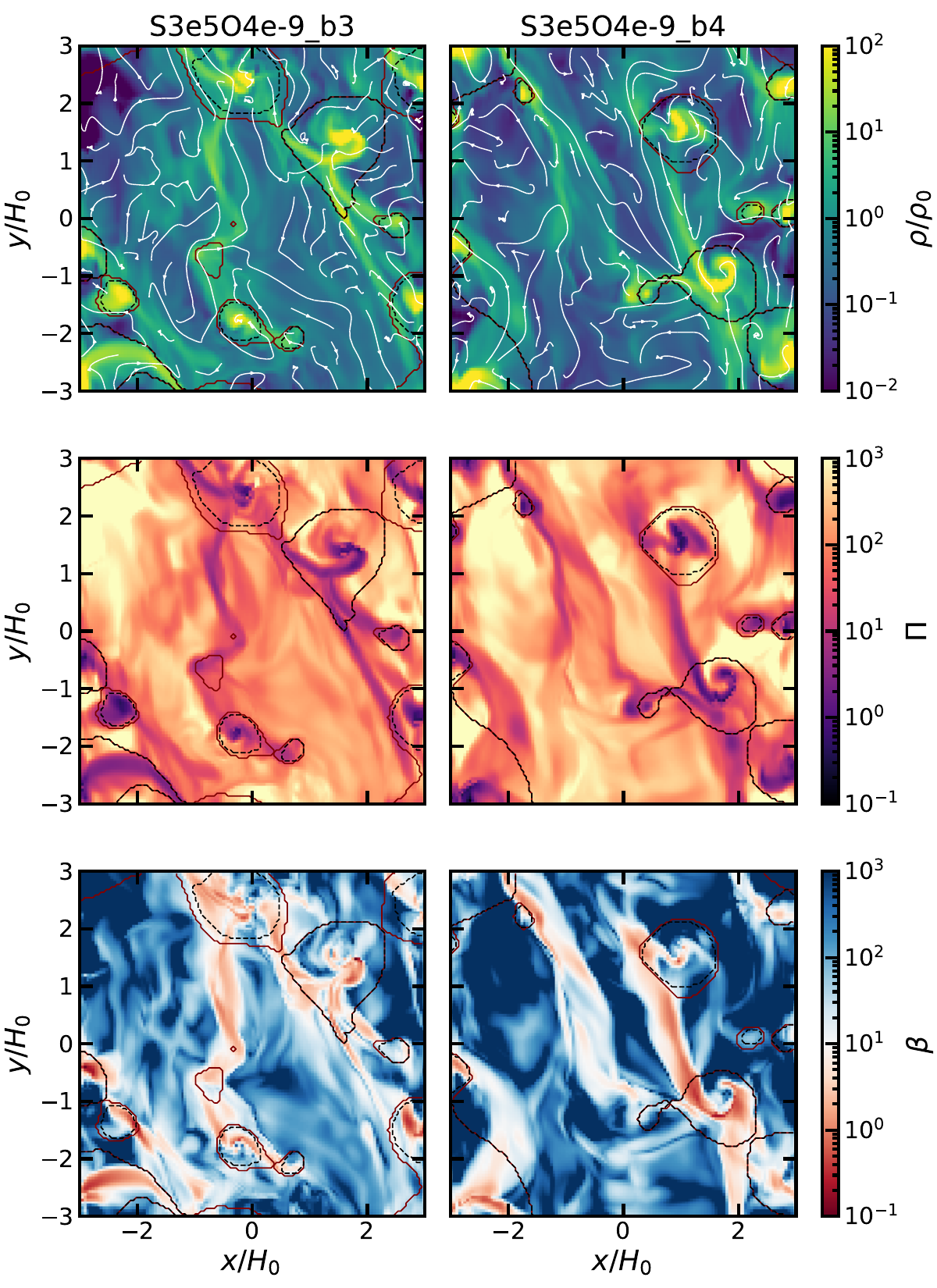}
\caption{Midplane snapshots of density $\rho/\rho_0$ (top), radiation-to-gas pressure ratio $\Pi$ (middle), and plasma beta $\beta$ (bottom) for the radiation-pressure-dominated runs \texttt{S3e5O4e-9\_b3} (left) and \texttt{S3e5O4e-9\_b4} (right) after the onset of fragmentation. White streamlines in the top panels show the in-plane magnetic field $(B_x,B_y)$ at the midplane. Coordinates are normalized by the initial scale height $H_0$. Brown solid contours mark the boundaries of hierarchical binding parents (HBPs), while dashed black contours indicate gravitationally bound regions (HBRs; see \S~\ref{sec:bound}). Most HBPs contain bound central regions that continue to contract, whereas some overdensities remain unbound despite hosting local potential minima.}
\label{fig:rad_snapshot}
\end{figure*}

To examine clump properties, 
we show in \autoref{fig:rad_snapshot} a representative snapshot of the midplane density for the radiation-pressure-dominated, fragmenting runs after the formation of bound objects. 
Brown contours mark the boundaries of hierarchical binding parents (HBPs), 
while dashed black contours indicate gravitationally bound regions (see \S \ref{sec:bound}). 
Overlap between these contours indicates that an entire HBP is bound.
Most HBPs contain a bound central region that occupies a substantial fraction of the structure, 
with strong overdensity at the center and is expected to continue contracting. 
Some overdensities that satisfy the HBP criterion host local minima in the gravitational potential, 
but the potential wells are not sufficiently deep relative to their thermal and magnetic energy to be formally bound.
At very high overdensities, 
the dynamical timescales become extremely short, severely restricting the numerical timestep. 
As a result, we do not follow the long-term evolution of these clumps. 
We therefore cannot determine whether subsequent mergers and accretion lead to larger final masses (e.g., 
reaching conditions for nuclear fusion or core collapse) or whether the initially collapsing structures fragment into lower-mass objects. 
Nevertheless, 
these snapshots still allow for rough estimates of their initial properties.

\begin{figure*}[htbp]
\centering
\includegraphics[width=0.9\textwidth,clip=true]{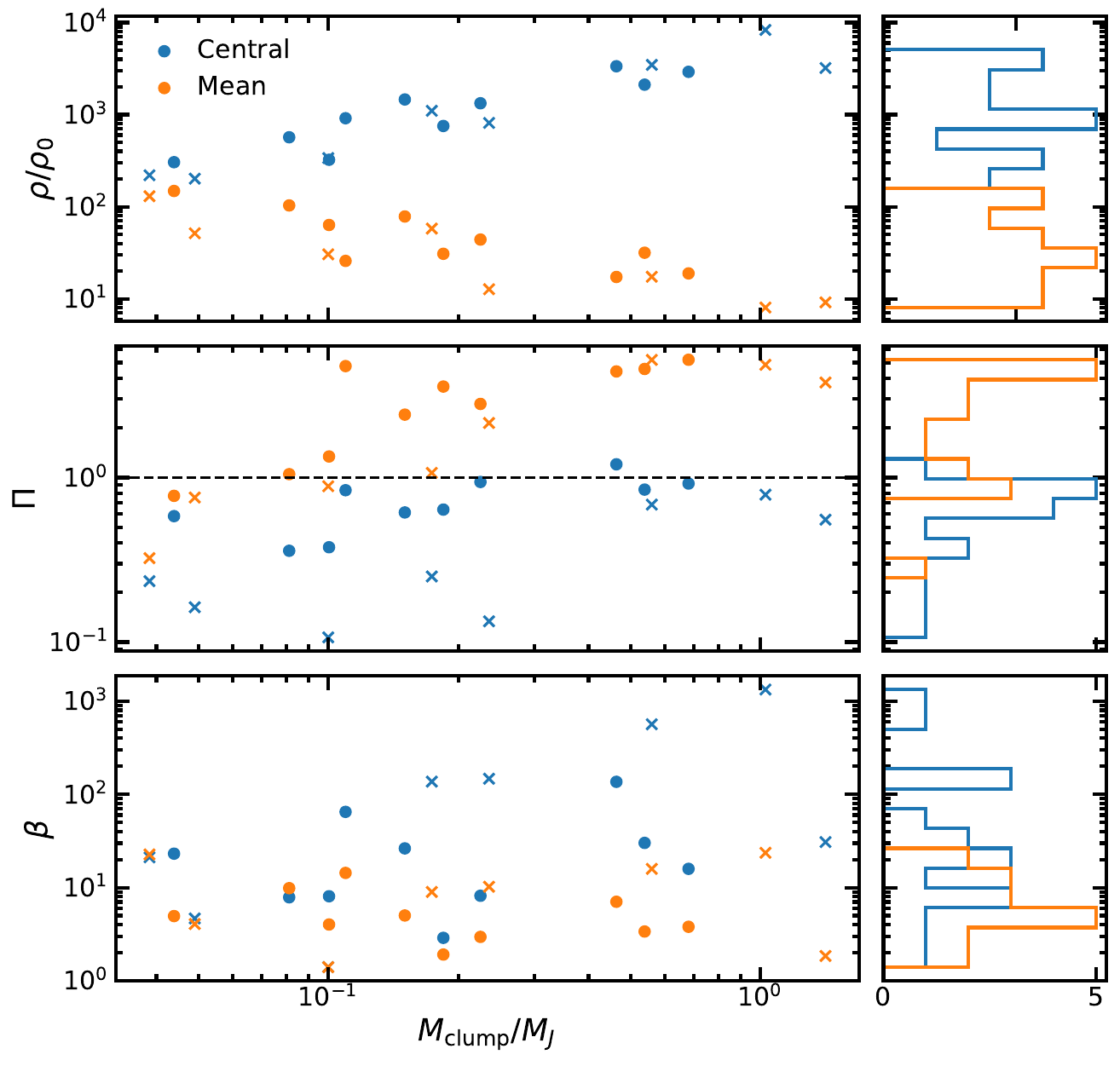}
\caption{Clump statistics for runs \texttt{S3e5O4e-9\_b4} (crosses) and \texttt{S3e5O4e-9\_b3} (circles). From top to bottom, we show the peak (central; blue) and mean (orange) densities normalized by $\rho_0$, the radiation-to-gas pressure ratio $\Pi$, and the plasma beta $\beta$, all as functions of clump mass in units of $M_J$. Central values are evaluated at the cell of maximum density within each clump, while mean values are volume-averaged over the bound region. The dashed line in the middle panel marks $\Pi=1$.}
\label{fig:clumps_raddominated}
\end{figure*}

In Figure \ref{fig:clumps_raddominated}, we plot the mean and peak (or ``central") density of each clump versus their masses measured in 
\begin{multline}
M_{\rm J} \equiv \frac{\pi^2 G^2\Sigma^3}{\Omega^4}
\simeq 2000 M_\odot \\
\times\left(\frac{\Sigma}{3\times10^5\,\mathrm{g\,cm^{-2}}}\right)^3
\left(\frac{\Omega}{4\times10^{-9}\,\mathrm{s^{-1}}}\right)^{-4}
\end{multline}
which is the typical clump mass in the absence of magnetic fields. The dots correspond to run \texttt{S3e5O4e-9\_b3} and the crosses are from run \texttt{S3e5O4e-9\_b4}. 
The most massive clumps are of order $\sim M_J$, 
although there is a tail of bound regions extending to lower masses. 
The presence of these lower-mass fragments 
{unstable modes that extend to the gas pressure Jeans scale due to radiative diffusion \citep{Thompson2008, Chen2023}, which implies the smallest clumps formed in regions with strong local cooling can be as small as $M_{J,\rm gas}\sim M_{\rm J}/\Pi$.} 
The smallest structures also tend to have $\rho_{\rm mean} \sim \rho_{\rm max}$ 
due to their limited spatial extent. 
It remains unclear whether subsequent evolution will grow their masses and increase their density contrast, potentially bringing them into alignment with the $\sim M_J$ population.
Notably, 
the total mass contained in bound regions already reaches $\sim40\%$ and $50\%$ of total mass in the box for run \texttt{S3e5O4e-9\_b3} and \texttt{S3e5O4e-9\_b4} respectively, 
suggesting that the available reservoir for further accretion may be limited unless mergers occur frequently.

We also show in Figure \ref{fig:clumps_raddominated} the mean values of $\Pi$ and $\beta$ (orange), 
along with their central values evaluated at the cell of maximum density (blue), to characterize the pressure partition within clumps.
The radiation pressure fraction is significantly reduced from its initial value, $\Pi(t=0)\approx 100$, due to cooling and contraction near the midplane (see Figure \ref{fig:vertical_raddominatedRMHD}), 
and falls to $\lesssim 1$ at clump centers. 
As gas pressure becomes dominant, further contraction is expected to be moderated.
Run \texttt{S3e5O4e-9\_b3} generally exhibits higher $\Pi$ values, likely due to enhanced small-scale heating from MRI-driven turbulence that partially offsets net cooling. 
Although this run also tends to have smaller $\beta$ values, 
indicating that the clumps are more strongly threaded by magnetic fields, the fact that both the mean and central values satisfy $\beta > 1$ implies that magnetic pressure remains subdominant within clumps in both cases.

Interestingly, \citet{Deng2021} use global MHD simulations of protoplanetary disks with a prescribed $\beta$-cooling law to follow fragmentation and the subsequent evolution of a small population of relatively isolated clumps. They find that gas rotation around these clumps twists and amplifies the ambient magnetic field, producing magnetically dominated envelopes. The midplane magnetic streamlines in the top panels of Figure~\ref{fig:rad_snapshot} show no comparably clear pattern of field wrapping around our clumps. One possible explanation is that coherent circulation is disrupted by clump--clump interactions, tidal forcing from nearby fragments, and the surrounding gravito-turbulence. At the Hill radius $R_{\rm H}$, the characteristic dynamical time is of order $\Omega^{-1}$, comparable to the winding time for circulation on that scale. A neighbour with separation $\sim R_{\rm H}$ could therefore perturb the flow before a coherent magnetic envelope develops. This may reflect a difference between relatively isolated fragments with shear-dominated fluid motion and a chaotic, many-fragment environment.

\section{Summary and Discussions}
\label{sec:SF_discussion}

In this paper, we performed RMHD shearing-box simulations with self-gravity to
investigate the interplay between gravitational instability (GI) and the
magnetorotational instability (MRI). We focused on moderately magnetized
regimes, $\beta_z = 10^3$ and $10^4$, which avoid strong outflows and
significant mass loss. Each disk is initialized at $Q_T > 1$ and allowed to
cool while gravito-turbulence and MRI-driven dynamo activity develop. We adopt
combinations of $(\Sigma, \Omega)$ corresponding to gas-pressure- and
radiation-pressure-dominated regimes, with comparable optical depths and
cooling timescales.

In the gas-pressure-dominated runs, the MRI efficiently generates Maxwell
stress, and the resulting heating can exceed that from gravito-turbulence. In
some cases it drives the disk to hotter states in which GI is suppressed. The
radiation-pressure-dominated runs, by contrast, still fragment. Because
radiation-dominated gas is highly compressible, MRI turbulence produces weaker
stresses for a given perturbation amplitude, so heating cannot offset cooling.
Reaching comparable stresses instead requires larger density fluctuations,
which directly promote the collapse of overdense regions. Magnetic pressure
remains subdominant within the resulting clumps. Their characteristic masses
are of order $M_J$, together with a population of lower-mass fragments. These
clumps may seed star formation even in magnetized AGN disks.

We have not explored the strongly magnetized regime in which the midplane
remains persistently magnetically dominated, which may require
$\beta_z \lesssim 10^2$. Our attempts to reach
quasi-steady states there were limited by strong magnetically driven outflows
and the resulting mass loss. 
Mass can be replenished by rescaling, as in
\citet{BaiStone2013,Salvesen2016,Tsung2025}, 
but no current scheme does so while preserving the energy
budget for RMHD systems, and that budget is central to interpreting our results. 
Controlled
long-term evolution in this regime therefore is numerically challenging.

An alternative framework is to initialize the disk in a magnetically dominated
state with $Q_T \gg 1$ 
and follow how radiative cooling drives its subsequent
contraction. 
\citet{Zhang2026,Chen2026} suggest that strongly magnetized states may be difficult to
maintain in disks that efficiently cool and would tend to collapse into a classic MRI-heated disk or 
reconnection-heated disk, 
either supported by thermal pressure, although more parameter survey is needed to probe this regime. 
These simulations need not include
self-gravity; 
they can instead isolate the interplay between radiative cooling
and the MRI in the presence of strong azimuthal flux or NVF,
particularly for $\Pi \gg 1$, 
a regime largely unexplored in the literature.
Establishing this baseline may be a prerequisite for assessing the role of GI in
strongly magnetized disks, and 
we defer it to future shearing-box studies
with improved treatments of mass and energy replenishment, 
or global
simulations 
which constrain additional degrees of freedom more naturally.

In circumstances where fragmentation does occur, 
an important parallel study is to follow the subsequent contraction of self-gravitating clumps into one or more stars towards the main sequence. This is not explored in our current simulations due to limitations of the \texttt{Athena++} self-gravity solver in shearing-box setups.
Such evolution may be more naturally captured with Lagrangian methods \citep{Ni2025,Ni2026}. 
Relevant works \citep{Chen2025,Chen2026c} applied spherical coordinates centered on an embedded star within an AGN disk, 
optionally including tidal potentials and background shear flow. 
These models focus on the accretion onto a single stellar object that is already well past the initial contraction phase, whose gravity can be treated as a point source. 
However, those setups offer significantly higher resolution toward the center 
and 
allows for coupling between radiative transfer and a spherical coordinate self-gravity solver \citep{Xu2021}, 
suggesting that it could also be applied to study the contraction process in future works.

\section*{Acknowledgement}
We thank Douglas Lin, Zhaohuan Zhu, Phil Armitage and Jim Stone for helpful discussions. Simulations in this work are performed with computational resources at the Flatiron Institute and Princeton Institute for Computational Science and Engineering (PICSciE).

\appendix
\section{Non-self-gravitating Analog of Radiation Pressure Dominant Run}
\label{app:nsg}

To our knowledge, stratified radiation-MHD simulations of
radiation-pressure-dominated disks with NVF remain limited.\footnote{A related calculation is presented by
\citet{Scepi2018}, who considered an NVF, gas-pressure-dominated disk
with flux-limited-diffusion radiation transport in the context of
dwarf novae. 
The relevant instability there is associated primarily
with the nonlinear temperature dependence of the opacity rather than
radiation-pressure support.}
ZNVF simulations, by contrast, exhibit the
classical radiation-pressure thermal instability
\citep{Jiang2013}: 
sufficiently strong initial cooling can trigger
runaway contraction when turbulent heating increases more steeply
with pressure than radiative cooling. 
This outcome need not carry
over directly to NVF disks because their heating depends on both the
un-conserved thermodynamic pressure and the imposed NVF. 
We therefore performed a non-self-gravitating control
simulation 
initialized from the same radiation-pressure-dominated
state as our \texttt{S3e5O4e9\_b3},
with $\Pi \sim 100$ and a nominal
$Q=1.5$, 
but with self-gravity disabled. 
The initial hydrostatic equilibrium follows a strict $n=3$ polytrope \citep{Chen2025,Chen2026,Chen2026c} and cannot be identical to the self-gravitating runs. Nevertheless,
the purpose of this experiment
is not a systematic study of NVF thermal stability, 
but to determine
whether the corresponding disk undergoes prompt thermal collapse in
the absence of self-gravity. 

\begin{figure}
  \centering
\includegraphics[width=0.5\textwidth]{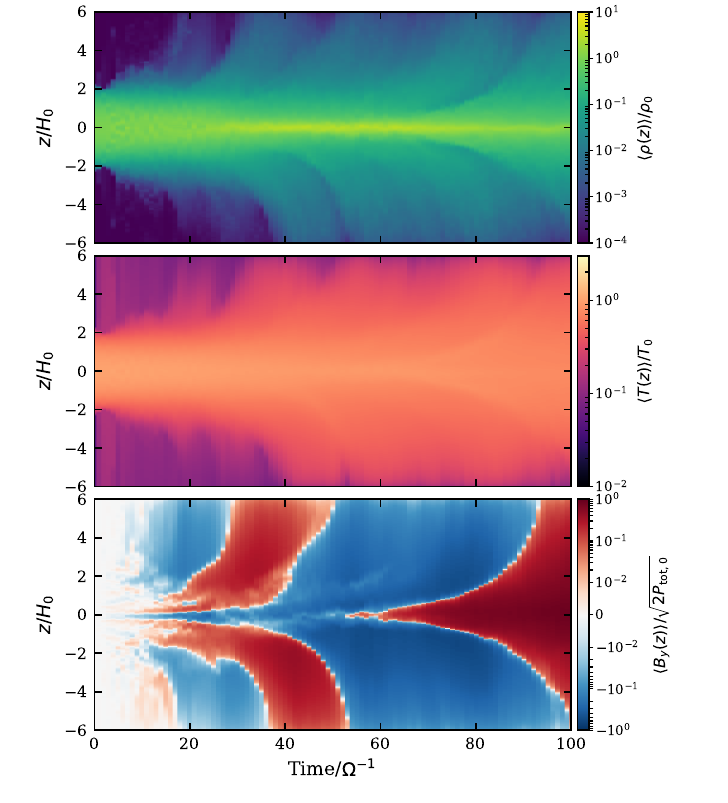}
  \caption{Time evolution of the horizontally averaged vertical structure in the non-self-gravitating NVF control simulation. 
  From top to bottom, the panels show density, temperature, and azimuthal magnetic field, normalized by their initial midplane reference values. 
  The disk eventually expands. }
  \label{fig:nsg-spacetime}
\end{figure}

As we show in Figure~\ref{fig:nsg-spacetime}, the control run does not collapse. 
The midplane density initially
increases mildly and reaches $\langle\rho\rangle_{\rm mid}\sim 2\rho_0$
at $t\sim 50\,\Omega^{-1}$, after which the disk expands and the
midplane density decreases monotonically. Towards the end of the run at 100$\Omega^{-1}$, the density scale height
has doubled. 
Angular-momentum
transport and heating is predominantly magnetic: averaging over
$t\Omega>50$, 
for example, the time-averaged stresses are
$\alpha_M\simeq0.1$ and $\alpha_R\simeq 0.02$. Thus, unlike the corresponding ZNVF radiation pressure dominant cases ($\alpha\sim 0.02$, therefore runaway cooling is expected to occur 
with initial $\tau_{\rm cool}\lesssim 50$ until $\Pi$ cools to order-unity), 
MRI heating
is sufficient to prevent an thermal collapse and ultimately
drives vertical expansion.

\begin{figure*}
  \centering
  \includegraphics[width=1.0\textwidth]
    {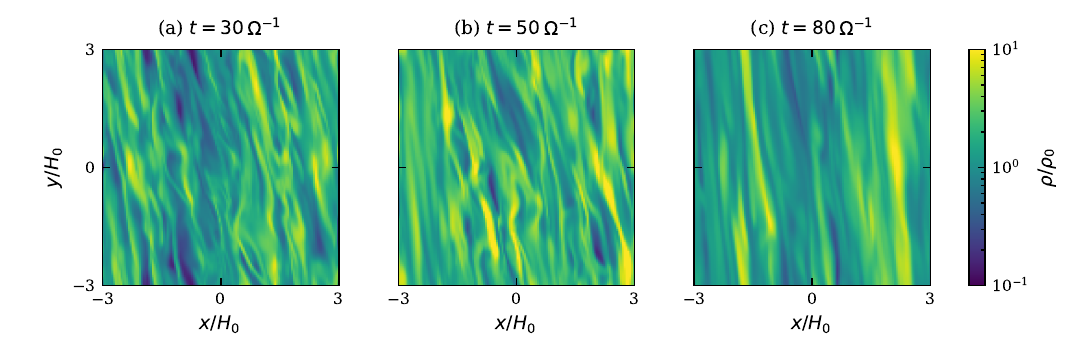}
  \caption{Midplane density at
  $t=30$, $50$, and $80\,\Omega^{-1}$ in the
  non-self-gravitating NVF control simulation. 
  All panels use the
  same logarithmic color scale.}
  \label{fig:nsg-snapshots}
\end{figure*}

The late-time evolution should nevertheless be interpreted
cautiously. By $t=100\,\Omega^{-1}$, approximately $25\%$ of the
initial mass has left the domain, the midplane has become strongly
magnetized, and the characteristic structures approach the radial
box scale (Figure~\ref{fig:nsg-snapshots}). 
Extending the calculation much further would
therefore require a larger domain and better control of the vertical
mass flux. 
The density field also remains highly intermittent: across all snapshots after MRI has developed, 
regardless of the non-axisymmetry of the 
characteristic modes, the midplane density dispersion is consistently high 
$\delta \rho_{\rm mid} \sim 2.2 \rho_0$, and the
high-density tail easily reach $\rho_{\max}\simeq10\rho_0$.
These overdensities would be especially susceptible to
self-gravitating collapse, 
corroborating our argument in \S \ref{sec:radresults}.


\bibliography{sample701_new}{}
\bibliographystyle{aasjournalv7}

\end{CJK*}
\end{document}